\documentclass[12pt]{article}
\usepackage[dvips]{graphicx,color}
\usepackage{wrapfig}
\begin{document}

%
\begin{flushright}
OU-HET-591, November, 2007 \ \ \\
\end{flushright}
\vspace{-3mm}

\begin{center}
{\bfseries COMPARATIVE ANALYSIS OF TRANSVERSITIES AND LONGITUDINALLY
POLARIZED DISTRIBUTIONS OF THE NUCLEON}

\vskip 5mm
M. Wakamatsu$^{\dag}$

\vskip 5mm
{\small
(1) {\it
Department of Physics, Faculty of Science, Osaka University,\\
Toyonaka, Osaka 560-0043, JAPAN
}\\
$\dag$ {\it
E-mail: wakamatu@phys.sci.osaka-u.ac.jp
}}
\end{center}

\vskip 1mm
\begin{abstract}
We carry out a comparative analysis of the transversities and the
longitudinally polarized parton distribution functions in light
of the first empirical extraction of the transversity distributions
recently done by Anselmino et al. 
It is shown that the precise determination of the isoscalar tensor
charge, which is defined as the 1st moment of the isoscalar
combination of the transversity distributions, is of fundamental
importance for clarifying the internal spin structure function
of the nucleon. 
\end{abstract}

\vskip 3mm

As is well known, the transversity is one of the three fundamental
parton distribution functions (PDFs) with the lowest twist 2.
Different from the other two, i.e. more familiar unpolarized PDF
and the longitudinally polarized PDF, its chiral-odd nature prevents
us from extracting it directly through the standard inclusive
deep-inelastic-scattering measurements \cite{JJ92},\cite{BDR02}.
For this reason, we have had
little empirical information on it until recently.
Very recently, however, Anselmino et al. succeeded to get a first empirical
information on the transversities \cite{ABDKMPT07} from the combined
global analysis of the azimuthal asymmetries in semi-inclusive DIS
scatterings measured by HERMES and COMPASS groups
\cite{HERMES05},\cite{COMPASS07}, and those in
$e^+ e^- \rightarrow h_1 h_2 X$ processes by the Belle
Collaboration \cite{Belle06}.
Their main observation for the transversities can be summarized as
follows. First, the $u$-quark transversity is positive and $d$-quark one
is negative with the magnitude of $\Delta_T u(x)$ being much larger than
that of $\Delta_T d(x)$. 
Second, both of $\Delta_T u(x)$ and $\Delta_T d(x)$ are
significantly smaller than the Soffer bound \cite{Soffer95}. The 2nd
observation is only natural, since the magnitudes of unpolarized
PDFs are generally much larger than the polarized PDFs.
In our opinion, what is more interesting from the physical viewpoint
is the comparison of the transversities with the longitudinally
polarized PDFs. This comparative analysis of the two fundamental PDFs
is the main purpose of my present talk \cite{Wakamatsu07T}.

Before going into the comparative analysis of the transversities and
the longitudinally polarized PDFs, it would be useful to give an overview
of new measurements of the longitudinally polarized
PDFs, especially in the flavor singlet channel related to the
nucleon spin problem. Recently, the COMPASS and HERMES groups
carried out high-statistics measurements of the longitudinal spin
structure function of the deuteron, thereby having succeeded to significantly
reduce the error bars of $\Delta \Sigma$, the net quark spin contribution to
the nucleon spin \cite{COMPASS05D}-\nocite{COMPASS06D}\cite{HERMES06D}. 

\begin{figure}[h]
\begin{center}
\begin{minipage}[l]{7.0cm}
\includegraphics[width=70mm,height=70mm]{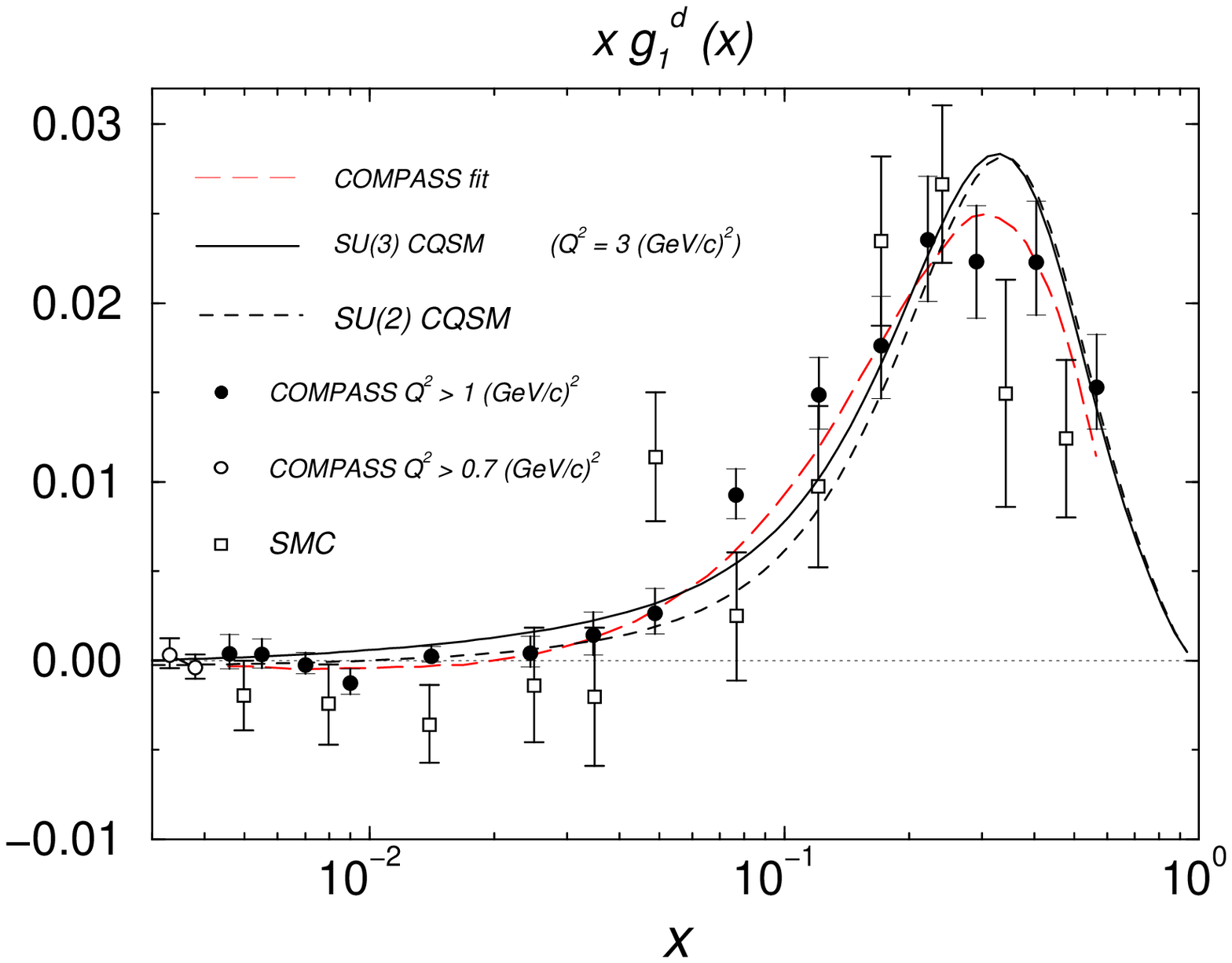}
\caption{\small The predictions of the $SU(2)$ and $SU(3)$
  CQSM in   comparison with the new COMPASS data for $x \,g_1^d (x)$
  (filled circles) and their NLO QCD fits (long-dashed curve).
  The old SMC data \cite{SMC98} are also shown by open squares.}
\vspace{-4mm}
\label{Fig1}
\end{minipage}
\begin{minipage}[c]{2.0cm}
\end{minipage}
\begin{minipage}[r]{8.0cm}
\includegraphics[width=70mm,height=70mm]{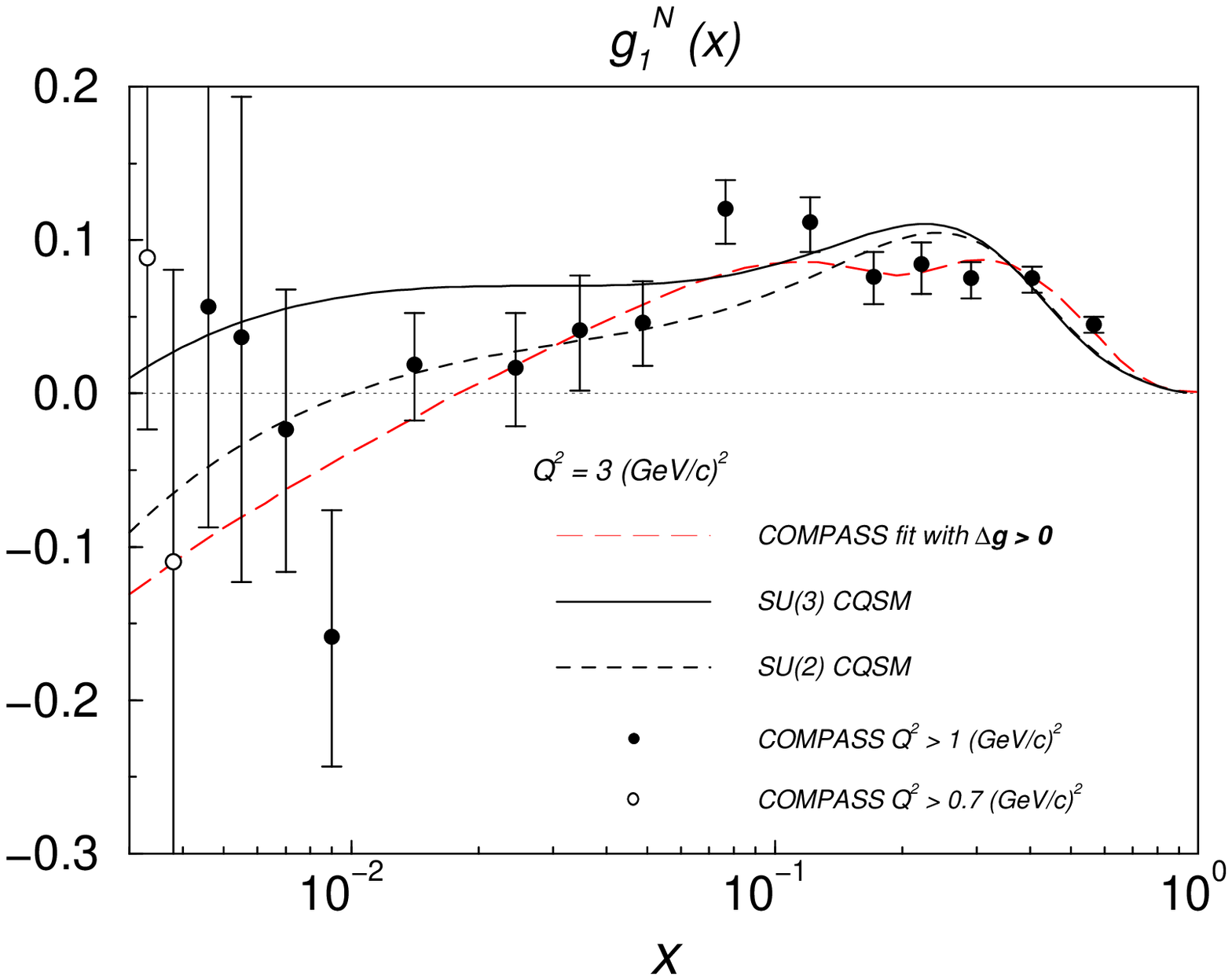}
\vspace{0mm}
\caption{\small The predictions of the $SU(2)$ and $SU(3)$
  CQSM in comparison with the new COMPASS data for $g_1^N (x)$
  (filled circles) and their NLO QCD fits (long-dashed curve).}
\vspace{4mm}
\label{Fig2}
\end{minipage}
\end{center}
\end{figure}

As pointed out in \cite{Wakam07C}, these new results for the deuteron
spin structure function is remarkably close to our theoretical
predictions given some years ago based on the chiral quark soliton
model (CQSM) \cite{Wakam07C},\cite{Wakamatsu03}. (See also
\cite{DPPPW96}-\nocite{DPPPW97}\nocite{WGR96}\cite{WW00}.)
Fig.\ref{Fig1} show the comparison between our predictions for
$x \,g_1^d (x,Q^2)$ given several years ago and the new COMPASS
data \cite{COMPASS05D} (the filled circles) together with the
old SMC data \cite{SMC98} (the open squares). 
The solid and dashed curves respectively stand for the predictions
of the flavor $SU(3)$ and $SU(2)$ CQSM evolved to the energy scale
$Q^2 = 3 \,\mbox{GeV}^2$, which is the average energy scale of the
new COMPASS measurement. The long-dashed curve shown for reference
is the next-to-leading order QCD fit by the COMPASS
group \cite{COMPASS06D}.
As one can see, the new COMPASS data show a considerable deviation
from the old SMC data in the small $x$ region. One finds that
the predictions of the CQSM are consistent with the new
COMPASS data especially in the small $x$ region.
This tendency can more clearly be seen in comparison of $g_1^N (x)
\equiv g_1^d (x) / (1 - \frac{3}{2} \,\omega_D)$ illustrated in
Fig.\ref{Fig2}.
The filled circles here represent the new
COMPASS data for $g_1^N (x)$, while the long-dashed curve is the
result of the next-to-leading order QCD fit by the COMPASS
group \cite{COMPASS06D}.
The predictions of the $SU(3)$ and $SU(2)$ CQSM are represented by the
solid and dashed curves, respectively. For the quantity $g_1^N (x)$,
the experimental uncertainties are still fairly large in
the small $x$ region. Still, one can say that the predictions of
the CQSM is qualitatively consistent with the new COMPASS data
as well as their QCD fit.

The COMPASS group also extracted the matrix element of the
flavor-singlet axial charge $a_0$ \cite{COMPASS06D},
which can be identified with
the net longitudinal quark polarization $\Delta \Sigma$ in the
$\overline{\rm MS}$ factorization scheme. Taking the value of
$a_8$ from the hyperon beta decay, under the assumption of
$SU(3)$ flavor symmetry, they extracted from the QCD fit of the
new COMPASS data for $g_1^d(x)$ the value of $\Delta \Sigma$ as
\begin{equation}
 \Delta \Sigma (Q^2 = 3 \,\mbox{GeV}^2)_{COMPASS} \ = \ 0.35 
 \ \pm \ 0.03 \,(stat.) \ \pm \ 0.05 \,(syst.) .
\end{equation}
On the other hand, the same quantity derived from the fits to all
$g_1$ data is a little smaller
\begin{equation}
 \Delta \Sigma (Q^2 = 3 \,\mbox{GeV}^2)_{COMPASS} \ = \ 0.30
 \ \pm \ 0.01 \,(stat.) \ \pm \ 0.02 \,(evol.) .
\end{equation}
A similar analysis was also reported by the HERMES
group \cite{HERMES06D}. Their result is
\begin{equation}
 \Delta \Sigma (Q^2 = 5 \,\mbox{GeV}^2)_{HERMESS} \ = \ 0.330
 \ \pm \ 0.011 \,(theor.) \ \pm \ 0.025 \,(exp.) \ \pm \ 0.028 \,(evol.) .
\end{equation}

\begin{wrapfigure}{R}{8cm}
\begin{center}
\vspace{-5mm}
\includegraphics[width=80mm]{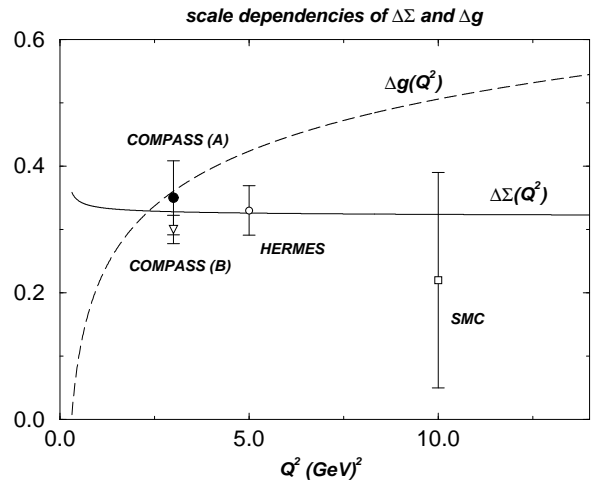}
\vspace{-10mm}
\end{center}
\caption{\small The scale dependencies of $\Delta \Sigma$ and
  $\Delta g$   predicted by the CQSM in combination with the NLO DGLAP
  equation   are compared with the recent QCD fits by the COMPASS group 
  (filled circle and open triangle) and by the
  HERMES group (open circle). The old SMC result is also
  shown by an open square.}
\label{Fig3}
\end{wrapfigure}

The results of the two groups for $\Delta \Sigma$ are mutually
consistent and seems to be larger than the previously known central
values \cite{SMC98}. We now compare these new results with
the prediction of the CQSM given in our previous
papers \cite{WK99},\cite{Wakamatsu03}.
Shown in Fig.\ref{Wakamatsu_fig3} are the prediction of the CQSM
for $\Delta \Sigma$ and $\Delta g$ as functions of the energy
scale $Q^2$.
They are obtained by solving the standard DGLAP equation at the NLO
with the prediction of the model as the initial condition given at the
scale $Q_{ini}^2 = 0.30 \,\mbox{GeV}^2 \simeq (600 \,\mbox{MeV})^2$.
Since the CQSM is an effective quark model, which contains no
gluon degrees of freedom, $\Delta g$ is simply assumed to be zero
at the initial scale. One sees that the new COMPASS and the
HERMES results for $\Delta \Sigma$ are surprisingly close to
the prediction of the CQSM. Also interesting is the longitudinal
gluon polarization $\Delta g$. In spite that we have assumed
that $\Delta g$ is zero at the starting energy, it grows rapidly
with increasing $Q^2$. As pointed out in \cite{Cheng96},
the growth of the gluon polarization with $Q^2$ can be traced
back to the positive
sign of the anomalous dimension $\gamma^{(0)1}_{qg}$. The positivity
of this quantity dictates that the polarized quark is preferred to
radiate a gluon with helicity parallel to the quark polarization.
Since the net quark spin component in the proton is positive,
it follows that $\Delta g > 0$ at least for the gluon perturbatively
emitted from quarks. The growth rate of $\Delta g$ is so fast
especially in the relatively small $Q^2$ region that its magnitude
reaches around $(0.3 - 0.4)$ already at $Q^2 = 3 \,\mbox{GeV}^2$, which
may be compared with the estimate given by the COMPASS group :
\begin{equation}
 \Delta g (Q^2 = 3 \,\mbox{GeV}^2)_{COMPASS} \ \simeq \ 
 (0.2 - 0.3).
\end{equation}

Now that we have convinced that the CQSM reproduces very well the
longitudinally polarized PDFs of the nucleon and the deuteron,
we return to the main topic of this talk, i.e. the
difference of the longitudinally
polarized PDFs and the transversities. First, I recall that the
most important quantities characterizing these PDFs are their
1st moments, known as the axial and tensor charges.
Next, I emphasize that the understanding of isospin dependencies is
crucially important to disentangle the
{\it nonperturbative chiral dynamics} of QCD hidden in the PDFs.
Neglecting the strange quark degrees of freedom,
for simplicity, there exist two independent combinations. the
isoscalar and isovector combinations for both of the axial and tensor
charges. 

Let us first recall some basic facts about the axial and
tensor charges. The difference of the axial and tensor charges is
of purely relativistic nature \cite{JJ92}.
In fact, in the naive quark model or
the nonrelativistic quark model, there is no difference between
the axial and tensor charges, that is, the isovector axial and tensor
charges are both $5/3$, while the isoscalar axial and tensor charges
are both unity : 
\begin{equation}
 g_A^{(I=1)} \ = \ g_T^{(I=1)} \ = \ \frac{5}{3}, \ \ \ \ \ 
 g_A^{(I=0)} \ = \ g_T^{(I=0)} \ = \ 1.
\end{equation}
On the other hand, in the familiar MIT bag model,
which is nothing but the valence quark model with the relativistic
kinematics, an important difference appear between the axial and tensor
charges due to the presence of the lower component of the ground state
wave function $g(r)$ as
\begin{eqnarray}
 g_A^{(I = 0)} &=& \,1 \cdot \int \,
 \left(f^2 - \frac{1}{3} \,g^2 \right) \,r^2 \,d r, \ \ \ \ \ 
 g_A^{(I = 1)} \ = \ \frac{5}{3} \cdot \int \,
 \left(f^2 - \frac{1}{3} \,g^2 \right) \,r^2 \,d r, \\
 g_T^{(I = 0)} &=& \,1 \cdot \int \,
 \left(f^2 + \frac{1}{3} \,g^2 \right) \,r^2 \,d r, \ \ \ \ \ 
 g_T^{(I = 1)} \ = \ \frac{5}{3} \cdot \int \,
 \left(f^2 + \frac{1}{3} \,g^2 \right) \,r^2 \,d r.
\end{eqnarray}
Nevertheless, an important observation is that the {\it ratio} of
the isoscalar to isovector charge is just {\it common} for the axial and
tensor charges, i.e. they are three fifth in both of the NQM and the MIT
bag model : 
\begin{equation}
 \frac{g_A^{(I=0)}}{g_A^{(I=1)}} \ = \ 
 \frac{g_T^{(I=0)}}{g_T^{(I=1)}} \ = \ \frac{3}{5}.
\end{equation}
Most probably, this feature is related to a common shortcoming
of these models, that is, the lack of the spontaneous chiral symmetry
breaking mechanism.
One can convince it by comparing the predictions of the
MIT bag model with those of the CQSM, which
is an effective model of QCD taking account of the effect of
spontaneous chiral symmetry breaking in a maximal way.

\vspace{5mm}
\begin{table}[h]
\begin{center}
\begin{tabular}{cccc}
\hline
 & \ \ MIT bag \ \ & \ \ CQSM \ \ & \ \ Experiment \ \ 
\\ \hline \hline
$g_A^{(I=1)}$ & 1.06 & 1.31 & 1.267 \ (\mbox{scale independent}) \\
$g_A^{(I=0)}$ & 0.64 & 0.35 & \ \ 
0.330 $\pm$ 0.040 \ ($Q^2 = 5 \mbox{GeV}^2$)
\ \  \\
$g_T^{(I=1)}$ & 1.34 & 1.21 &  \\
$g_T^{(I=0)}$ & 0.88 & 0.68 &  \\
\hline
$g_A^{(I=0)}/g_A^{(I=1)}$ & 0.60 & 0.27 & 
$\sim $ 0.26 \ ($Q^2 = 5 \mbox{GeV}^2$) \\
$g_T^{(I=0)}/g_T^{(I=1)}$ & 0.60 & 0.56 &
\\ \hline
\end{tabular}
\vspace{4mm}
\caption{\small The predictions of the MIT bag model and of CQSM for
the axial and tensor charges in comparison with the empirical
information.}
\end{center}
\end{table}

As mentioned, in the MIT bag model, the ratio of the
isoscalar and isovector axial charges and also the ratio of
isoscalar and isovector tensor charges are both exactly 0.6.
On the other hand, the CQSM predicts that the ratio of the
axial charges is much smaller than that of the tensor charges.
This comes from the fact that the CQSM predicts very small
isoscalar axial charge just consistent with the EMC observation,
while its prediction for the isoscalar tensor charge is not
extremely different from the prediction of other low energy
effective models including the MIT bag model.

In any case, the predictions of the CQSM for the axial and tensor
charges can roughly be summarized as follows.
The isovector tensor and axial charges have the same order of magnitudes,
while the isoscalar tensor charge is not so small as the isoscalar
axial charge. From this analysis, we immediately expect the following
qualitative features for the transversity and the longitudinally polarized
PDFs. The isovector transversity distribution and the isovector
longitudinally polarized distribution would have the same order of
magnitude, while the isoscalar $\Delta_T q(x)$ is much larger than the
isoscalar $\Delta q(x)$, i.e.
\begin{equation}
 \Delta q^{(I = 0)} (x) \ \ll \ \Delta_T q^{(I = 0)} (x), \ \ \ \ 
 \Delta q^{(I = 1)} (x) \ \simeq \ \Delta_T q^{(I = 1)} (x) .
\end{equation}
In other words, we would expect the magnitude
of $d$-quark transversity is much smaller than that of $d$-quark
longitudinally polarized PDF :
\begin{equation}
 | \Delta_T d(x) | \ \ll \ | \Delta d(x) | .
\end{equation}

\begin{wrapfigure}{R}{8cm}
\begin{center}
\vspace{-15mm}
\includegraphics[width=85mm,height=70mm]{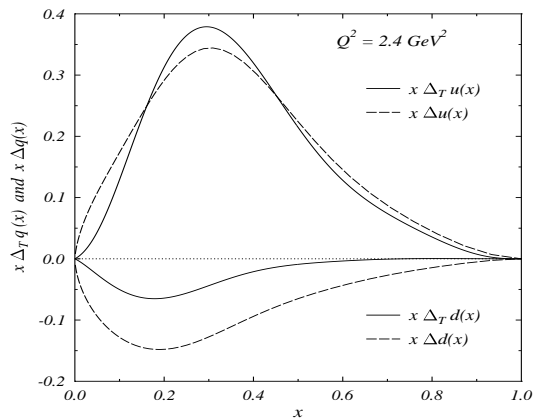}
\vspace{-15mm}
\end{center}
\caption{\small The predictions of the flavor $SU(2)$
  CQSM for the transversities (solid curves) and the longitudinally
  polarized  distribution functions (dashed curves) for the $u$- and
  $d$-quarks evolved to $Q^2 = 2.4 \,\mbox{GeV}$.}
\label{Fig4}
\end{wrapfigure}

To make the argument more quantitative, we compare in 
Fig.\ref{Wakamatsu_fig4} the CQSM predictions for the transversities and the
longitudinally polarized PDFs. Here, the model predictions are evolved to the energy scale of $Q^2 = 2.4 \,\mbox{GeV}^2$, for later convenience.
One can confirm that the magnitudes of the $u$-quark transversities
and the $u$-quark longitudinally polarized PDF are roughly the same,
whereas the magnitude of $d$-quark transversity is roughly a
{\it factor of two} smaller than that of the $d$-quark longitudinally
polarized PDF.

Now, I compare in Fig.\ref{Wakamatsu_fig5} the CQSM predictions
for the transversities with
the recently obtained global fit by Anselmino et al.~\cite{ABDKMPT07}.
As one sees, the uncertainties of the global fit are still quite large.
Still, a remarkable feature of the transversity distributions
seems to be already seen in their fit.
A common feature of the CQSM prediction and their global fit is that
the ratio $\Delta_T d(x)/\Delta d(x)$ is very small.
As a general trend, however, the magnitudes of the transversities
obtained by their global fit look fairly smaller than the
corresponding CQSM predictions.
In particular, the CQSM prediction for the $u$-quark transversity
appears to lie outside the upper limit of their fit.
We shall come back to this point later.

At this point, it would be useful to make some comments on the
calculation of transversities by Bochum group based on the same
CQSM \cite{SUPWPG01}. A main difference between our calculation
\cite{WK99}, \cite{Wakamatsu01} and theirs \cite{SUPWPG01}
resides in the isovector part of transversities
$\Delta_T q^{(I=1)}(x)$.
In their calculation, they included only the leading-order contribution
to this quantity, and neglected the subleading $1 / N_c$ correction,
while we have included the latter as well. This is because we
know that a similar $1 / N_c$ correction (or more concretely, the
1st-order rotational correction) is very important for resolving
the famous underestimation problem of some isovector observables,
like the isovector axial-charge and/or the isovector magnetic
moment of the nucleon, inherent in the hedgehog-type soliton
model \cite{WW93},\cite{CBGPPWW94}.
The neglect of this $1 / N_c$ correction would led to a similar
underestimation of the isovector tensor charge, thereby having a fear of
being lead to a misleading conclusion on the size of the
transversities. We emphasized that, to avoid such a danger, it is very
important to analyze the transversities and the longitudinally
polarized PDFs {\it simultaneously} within the same theoretical
framework.

\vspace{-5mm}
\begin{figure}[h]
\begin{center}
\begin{minipage}[l]{7.0cm}
\vspace{-5mm}
\includegraphics[width=80mm,height=70mm]{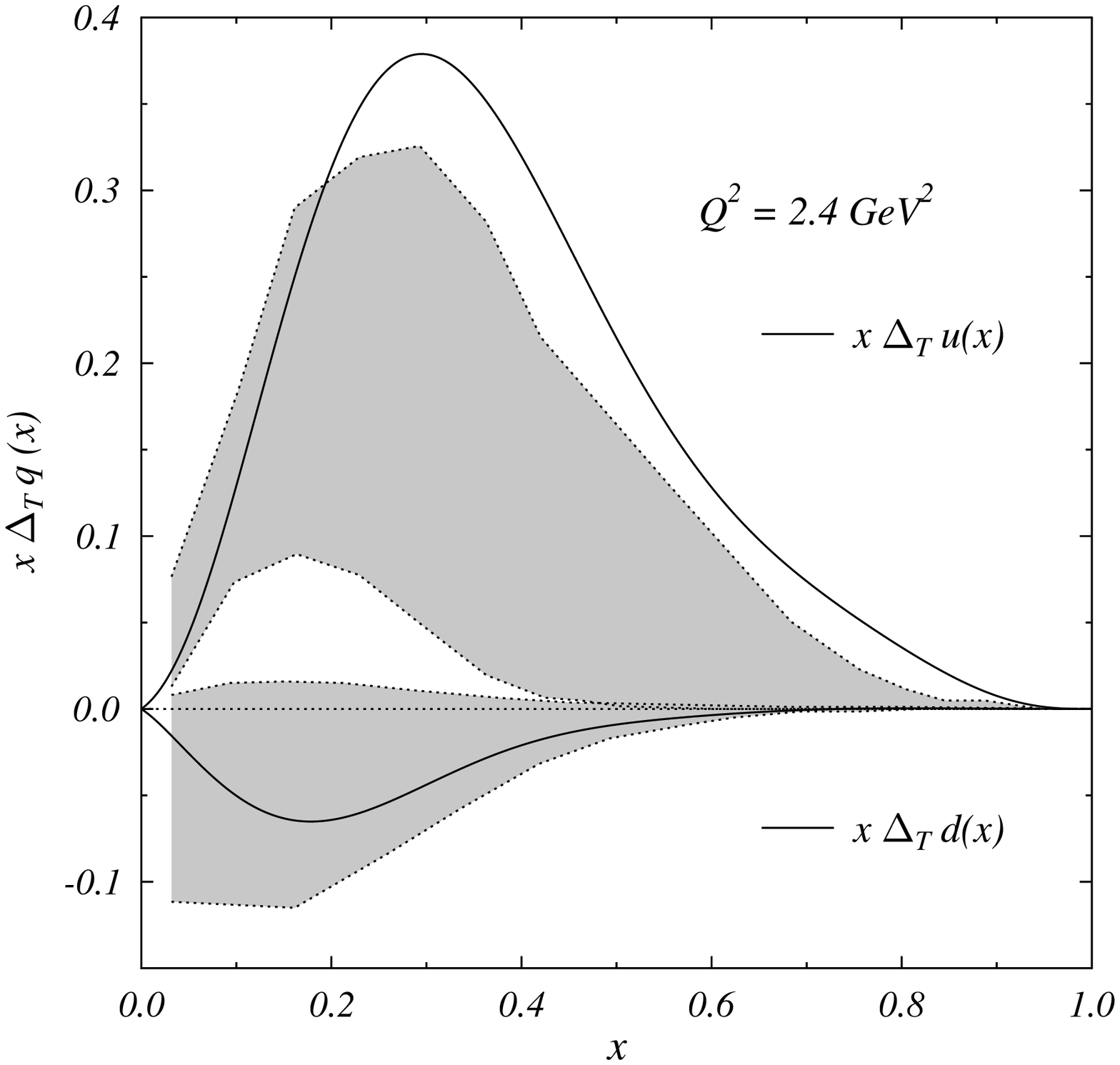}
\caption{\small The predictions of the flavor $SU(2)$ CQSM
  for the transversities (solid curves) in comparison with the
  global-fit of \cite{ABDKMPT07} (shaded areas).}
\label{Fig5}
\end{minipage}
\begin{minipage}[c]{2.0cm}
\end{minipage}
\begin{minipage}[r]{8.0cm}
\includegraphics[width=80mm,height=70mm]{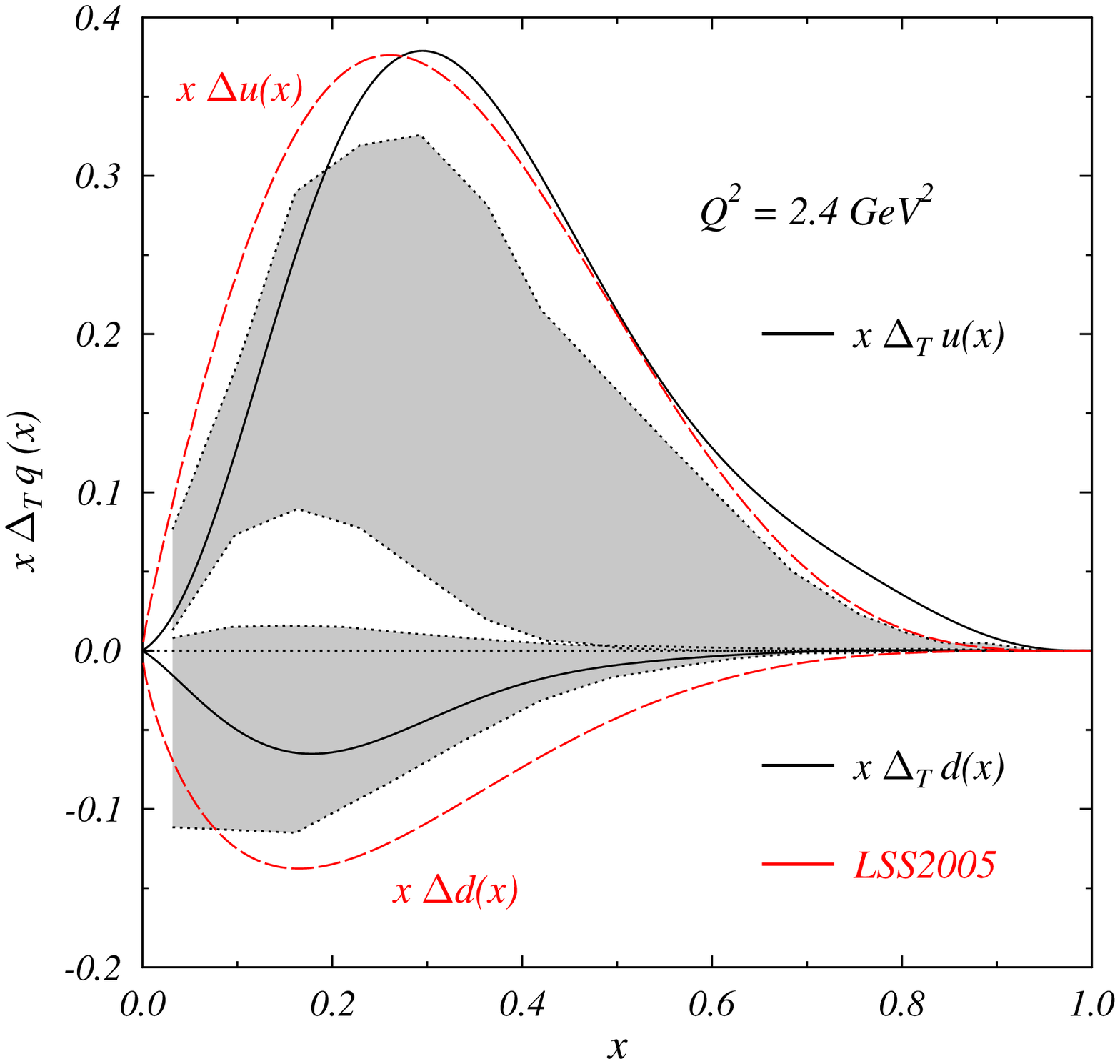}
\caption{\small The predictions of the flavor $SU(2)$ CQSM
  for the transversities (solid curves) in comparison with the
  LSS2005 fit \cite{LSS2005} of the longitudinally polarized
  $u$- and $d$-quark distributions.}
\label{Fig6}
\end{minipage}
\end{center}
\end{figure}

To see the difference with the longitudinally polarized PDFs,
we show in Fig.\ref{Wakamatsu_fig6} the LSS2005 fit for the
longitudinally polarized $u$- and $d$-quark
distributions \cite{LSS2005}.
One can confirm that the CQSM prediction for the $u$-quark
transversity has the same order of magnitude as that of the
LSS fit for the $u$-quark longitudinally polarized PDF, while
the CQSM prediction for the $d$-quark transversity is a factor of
two smaller than the LSS fit for the longitudinally polarized
PDF \cite{LSS2005}.

As already emphasized, the reason of this difference can be traced
back to the fact that the isoscalar tensor charge is not so small
as the isoscalar axial charge in the CQSM. Then, the next question is
why the CQSM predicts so small isoscalar axial charge.
First, I recall that in the standard $\overline{MS}$ scheme the
isoscalar axial charge can be identified with the net quark
polarization $\Delta \Sigma$. Within the framework of the CQSM,
we can prove the following nucleon spin sum rule, naturally saturated
by the quark fields alone \cite{WY91} :
\begin{equation}
 \frac{1}{2} \ = \ \frac{1}{2} \,\Delta \Sigma \ + \ L^Q .
\end{equation}
On the other hand, in accordance with
the physical nucleon picture of the model as a rotating hedgehog,
the CQSM predicts quite large quark OAM, which in turn dictates that
$\Delta \Sigma$ must be small \cite{WY91}.
As a matter of course, in real QCD, the correct nucleon spin sum rule
contains the gluon contributions as well :
\begin{equation}
 \frac{1}{2} \ = \ \frac{1}{2} \,\Delta \Sigma \ + \ L^Q \ + \ 
 \Delta g \ + \ L^g .
\end{equation}
However, all the recent investigations indicate that the $\Delta g$
is likely to be small at least in the relatively low energy scale.
Combining these observation, one must therefore conclude that
the sum of $L^Q$ and $L^g$ must be fairly large at low energy scale.

Our next question is then, "Is there any sum rule that constrains
the magnitudes of the isoscalar tensor charge ?
Here, one may remember the nucleon spin sum rule proposed by
Bakker, Leader and Trueman some years ago \cite{BLT04}, which
in fact contains the transversity distributions as
\begin{equation}
 \frac{1}{2} \ = \ \frac{1}{2} \,\sum_{a = q, \bar{q}} \,
 \int_0^1 \, \Delta_T q^a (x) \ + \ 
 \sum_{a = q, \bar{q}, g} \,\langle L_{s_T} \rangle^a ,
 \label{eq:BLT}
\end{equation}
where $L_{s_T}$ is the component of the orbital angular momentum
$\mbox{\boldmath $L$}$ along the transverse spin direction $s_T$.
Unfortunately, there are several peculiarities in the BLT sum rule.
First of all, it is not such a sum rule obtained as the 1st moment
of some parton distribution functions.
In fact, the r.h.s. of this sum rule does not correspond
to a nucleon matrix element of {\it local operator}.
In particular, the 1st term of this sum rule does not correspond
to the isoscalar tensor charge, because here the sum of the
quarks and antiquarks, not the difference, appear as
\begin{eqnarray}
 \sum_{a = q,\bar{q}} \, \int_0^1 \,\Delta_T \,q^a (x) \,dx
 &=& \int_0^1 \,\left\{ \,[ \Delta_T u(x) + \Delta_T d(x) ]
 \ + \ [ \Delta_T \bar{u} (x) + \Delta_T \bar{d}(x) ] \,\right\}
 \nonumber \\
 &\neq& \hspace{20mm} g_T^{(I=0)} \,.
\end{eqnarray}
Nonetheless, our analysis based on the CQSM indicates that
antiquark transversities are fairly small. This means that the 1st
term of the BLT sum rule may not be extremely different from
the tensor charge. Then, if the postulated inequality
between the isoscalar axial and tensor charges is in fact confirmed
experimentally, it would mean the following inequality, that is
the transverse OAM is much smaller than the longitudinal OAM :
\begin{equation}
 L_{s_T}^Q \ + \ L_{s_T}^g \ \ll \ L^Q \ + \ L^g .
\end{equation}

At this point, we come back to the discrepancy between the CQSM
predictions and the global fit by Anselmino et al.
We can estimate the magnitudes of
tensor charges from their central fit, under the
assumption that the antiquark contributions to them are negligible,
as justified by the CQSM. We then get the following values for the $u$-
and $d$-quark tensor charges, 
\begin{equation}
 \delta u \ \simeq \ 0.39, \ \ \ \ \Delta d \ \simeq \ - \,0.16,
\end{equation}
or for the isoscalar and the isovector tensor charges,
\begin{equation}
 g_T^{(I=0)} \ \simeq \ 0.23, \ \ \ \ g_T^{(I=1)} \ \simeq \ 0.55,
\end{equation}
at the energy scale $Q^2 \simeq 2.4 \,\mbox{GeV}^2$.
If they are evolved down to the low energy model
scale around $600 \,\mbox{MeV}$, we would obtain the following
numbers :
\begin{equation}
 \delta u \ \simeq \ 0.49, \ \ \ \ \Delta d \ \simeq \ - \,0.20,
\end{equation}
or
\begin{equation}
 g_T^{(I=0)} \ \simeq \ 0.28, \ \ \ \ g_T^{(I=1)} \ \simeq \ 0.69.
\end{equation}
We recall that all the theoretical estimates in the past,
based on the low energy models as well as the lattice QCD, predict
the isovector tensor charge between 1.0 and
1.5 \cite{HJ95}-\nocite{KPG96}\nocite{SS97}\nocite{PPB05}
\nocite{ADHK97}\cite{QCDSF05}.
At any rate, we emphasize that the transversities obtained by their
global fit correspond to fairly small magnitudes of tensor charges
as compared with the past theoretical estimates.

To sum up, we have carried out a comparative analysis of the
transversities and the longitudinally polarized PDFs in light of
the new global fit of transversities and the Collins fragmentation
functions carried out by Anselmino et al.
Their results, although with large uncertainties, already appears to
indicate a remarkable qualitative difference between transversities
and longitudinally polarized PDFs such that
$| \Delta_T d(x) / \Delta d(x) | \, \ll \, | \Delta_T u(x) / \Delta u(x) |$,
which is qualitatively consistent with the predictions of the CQSM.
I have emphasized that the cause of this feature can be traced back
to the relation $g_T^{(I=0)} \, \gg \, g_A^{(I=0)} \ = \ \Delta \Sigma$.
Further combining with the BLT sum rule, this indicates the
inequality, $L_{S_T}^Q \ + \ L_{S_T}^g \, \ll \, L^Q \ + \ L^g$,
i.e. the transverse OAM may be much smaller than the longitudinal
OAM. We are not sure whether this unique observation can be understood
as the dynamical effects of Lorentz boost or Melosh transformation.
Naturally, the global analysis carried out by Anselmino
et al. is just a 1st step for extracting transversities.
More complete understanding of the spin dependent fragmentation
mechanism is mandatory for getting more definite knowledge of the
transversities.
Also very desirable is some independent determination of transversities,
for example, through double transverse spin asymmetry in
Drell-Yan processes.
We hope that such near-future experiments will provide us with more
stringent constraint on the isovector as well as the isoscalar tensor
charges, thereby deepening our knowledge on the internal spin structure
function of the nucleon.

\vspace{6mm}
\begin{flushleft}
\begin{Large}
{\bf Acknowledgement}
\end{Large}
\end{flushleft}

\vspace{0mm}
This work is supported in part by a Grant-in-Aid for Scientific
Research for Ministry of Education, Culture, Sports, Science
and Technology, Japan (No.~C-16540253)


\end{document}